\documentstyle[12pt]{article}
\catcode`@=11
\newcount\@tempcntc
\def\@citex[#1]#2{\if@filesw\immediate\write\@auxout{\string\citation{#2}}\fi
  \@tempcnta\z@\@tempcntb\m@ne\def\@citea{}\@cite{\@for\@citeb:=#2\do
    {\@ifundefined
       {b@\@citeb}{\@citeo\@tempcntb\m@ne\@citea\def\@citea{,}{\bf ?}\@warning
       {Citation `\@citeb' on page \thepage \space undefined}}%
    {\setbox\z@\hbox{\global\@tempcntc0\csname b@\@citeb\endcsname\relax}%
     \ifnum\@tempcntc=\z@ \@citeo\@tempcntb\m@ne
       \@citea\def\@citea{,}\hbox{\csname b@\@citeb\endcsname}%
     \else
      \advance\@tempcntb\@ne
      \ifnum\@tempcntb=\@tempcntc
      \else\advance\@tempcntb\m@ne\@citeo
      \@tempcnta\@tempcntc\@tempcntb\@tempcntc\fi\fi}}\@citeo}{#1}}
\def\@citeo{\ifnum\@tempcnta>\@tempcntb\else\@citea\def\@citea{,}%
  \ifnum\@tempcnta=\@tempcntb\the\@tempcnta\else
   {\advance\@tempcnta\@ne\ifnum\@tempcnta=\@tempcntb \else \def\@citea{--}\fi
    \advance\@tempcnta\m@ne\the\@tempcnta\@citea\the\@tempcntb}\fi\fi}
\catcode`@=12
\def\barr{\begin{array}}
\def\earr{\end{array}}
\def\beq{\begin{equation}}
\def\eeq{\end{equation}}
\def\bea{\begin{eqnarray}}
\def\eea{\end{eqnarray}}
\def\bmath{\begin{displaymath}}
\def\emath{\end{displaymath}}
\def\bq{\begin{quote}}
\def\eq{\end{quote}}
\def\slash#1{\setbox0=\hbox{$#1$}#1\hskip-\wd0\hbox to\wd0{\hss\sl/\/\hss}}
\begin{document}

\begin{flushright}
RAL/93-074\\
October 1993
\end{flushright}

\begin{center}
{\bf{\large PROBING LEPTON NUMBER VIOLATION }}\\[0.3cm]
{\bf{\large VIA MAJORANA NEUTRINOS }}\\[0.3cm]
{\bf{\large AT HADRON SUPERCOLLIDERS }}\\[2.cm]
{\large A.~Datta}$^{a}$,
{\large M.~Guchait}$^{a}$
{\large and A.~Pilaftsis}$^{b}$\\[0.4cm]
$^{a}$ Physics Department, Jadavpur University, Calcutta 700 032,
India\\[0.3cm]
$^{b}$ Rutherford Appleton Laboratory, Chilton, Didcot, Oxon, OX11 0QX, UK.\\
\end{center}
\bigskip
\bigskip
\bigskip
\bigskip
\centerline {\bf ABSTRACT}

The possibility of discovering heavy Majorana neutrinos and lepton
number violation
via the like sign dilepton signal at hadron supercolliders is investigated.
The cross-sections for the production of these neutrinos singly as well
as in pairs are computed both in three and four generation scenarios
within the framework of the gauge group $SU(2)_L \otimes U(1)_Y$
and the dominant processes are identified. The suppression of the
Standard model background by suitable kinematical cuts is also discussed.

\newpage

\section*{I.~Introduction}

\indent

The present limits on the neutrino masses~\cite{PDG} reveal that even if
these masses are nonvanishing, they must be unnaturally small compared to
the corresponding quark or charged lepton masses.
An attractive solution to this
naturalness problem was inspired by the "see-saw" mechanism~\cite{YAN}
with the assumption that the neutrinos are Majorana fermions. In a simple
"see-saw" model with one generation of quarks and leptons, one obtains two
massive Majorana neutrinos, $\nu$ and $N$, having masses $m_\nu \simeq
m^2_D/m_M$ and $m_N \simeq m_M$. Thus if the Dirac mass of neutrinos $m_D$
is of the order of a typical quark or lepton mass and the Majorana mass
$m_M \gg m_D$, then $m_\nu$ can indeed be very small.

Originally the "see-saw" mechanism was contemplated in the context of
models (e.g.~grand unified theories ($GUT$'s) or left-right symmetric
models~\cite{GUT}) where the
scale $m_M$ is several orders of magnitude larger than the electroweak
scale. In such models the heavy neutrino mass is much beyond the reach of
the planned hadron supercolliders. Recently, however, simple extensions
of the Glashow-Salam-Weinberg standard model ($SM$) with Majorana mass
terms for the neutrinos have received much
attention~\cite{PROD,HP,DP,ZPC,APetal}.
These models based on the gauge group $SU(2)_L \otimes U(1)_Y$
and $m_M \sim 1$~TeV, predict heavy neutrinos well within the striking
ranges of $SSC$\footnote[1]{After completing our work, we became aware
of the disappointing news about the cancellation of the $SSC$ project.
However, our forthcoming analysis of the isolation of lepton-number violating
signals from the $SM$ background will show to be more relevant
for the $LHC$ collider.} and $LHC$. In particular, the observation of the
spectacular lepton-number violating decays of the heavy neutrinos via the
like sign dilepton ($LSD$) channel is of great experimental interest.

The coupling of the heavy neutrinos with $W$, $Z$ and Higgs ($H$) bosons
are, however, also naturally small. In an one generation "see-saw" model
the suppression factor $\xi = m_D/m_M$ turns out to be too small even for
$m_M\sim 1$~TeV, suppressing thereby the production cross-sections of these
neutrinos. It has, however, been pointed out that in realistic three
generation models, the neutrino masses are described by a $6\times 6$
matrix and the simple suppression as mentioned above may not
work~\cite{DP,ZPC,BW}.
But there are stringent experimental bounds on these
suppression factors from $LEP$ data
as well as from low energy experiments~\cite{LL,GBetal} which forces us to
accept
that this factor cannot be very large.

The purpose of the present work is to study the feasibility of observing
the $LSD$ signals at $LHC$ and $SSC$ by taking the most recent bounds on the
mixing angles into account. In Section II we estimate the cross sections
for the production of heavy Majorana neutrinos singly as well as in pairs,
via all possible channels. We then compute the cross section for the $LSD$
signal (using a parton level Monte Carlo calculation) for the dominant
process, which turns out to be $pp \to W^\ast \to l N X$, followed by the
lepton number violating decay of the heavy neutrino $N$. The kinematical
cuts required to suppress the $SM$ backgrounds arising primarily due to
heavy flavour production followed by cascade decays are also discussed.

The neutrino counting at $LEP$ strongly suggests that there are only three
light neutrinos within the framework of the $SM$. In an attempt to
demonstrate that the existence of a fourth family still remains a viable
possibility, it was shown that one can construct a simple extension of the
$SM$ with two naturally heavy Majorana neutrinos belonging to the fourth
generation~\cite{HP}. It was subsequently pointed out that the coupling of
these new neutrinos with $W$, $Z$ and $H$ are also naturally large~\cite{DP}.
As a result these neutrinos can be copiously produced at hadron colliders.
Production cross sections for heavy Majorana neutrino pairs were calculated
and they were found to be rather large~\cite{DP}. The number of $LSD$'s was
also estimated qualitatively.

In Section III we shall take up the question of producing the $LSD$ signal
in the context of the above four generation model in further details. The
$SM$ backgrounds and the relevant kinematical cuts required to suppress it
is also discussed. Our conclusions will be summarized in section IV.

\section*{II.~{\em LSD}'s in a three generation model }

\subsection*{II.1 The model}

\indent

Adopting the notation of Ref.~\cite{ZPC}, the relevant
interaction Lagrangian involving charged current is given by (summation
convention implied)
\beq
{\cal L}^{W-\nu_M-l}_{int} \ =\
-\ \frac{g_W}{\sqrt{2}} \ W^{-\mu} \Big[ \bar{l}_i \gamma_\mu \mbox{P}_L
(B_{l_ij} \nu_j \ +\ B_{l_i\alpha} N_\alpha ) \Big]\ +\ H.c.\ ,
\eeq 
where P$_L=(1-\gamma_5)/2$, $g_W$ is the coupling constant of $SU(2)_L$ and
$l$, $\nu$, $N$ and $W$ are respectively the lepton, light neutrino, heavy
neutrino and the W-boson field. The latin indices $i$, $j$, etc.$=
1,\dots,n_G$,
where $n_G$ denotes the number of generations, are used for
charged leptons and light  neutrinos,
while the greek indices $\alpha$, $\beta$,
etc.$=n_G+1, \dots, 2n_G$  indicate heavy Majorana neutrinos.
The neutral current interaction is given by
\bea
{\cal L}^{Z-\nu_M-\nu_M}_{int} &=& - \frac{g_W}{4\cos \theta_W}\  Z^\mu
\Bigg[ \bar{\nu}_i \gamma_\mu(i\mbox{Im}C_{ij}\ -\ \gamma_5 \mbox{Re}C_{ij})
\nu_j\nonumber\\
&+& \Big\{ \bar{\nu}_i\gamma_\mu (i\mbox{Im}C_{i\alpha}\ -\ \gamma_5
\mbox{Re}C_{i\alpha} ) N_\alpha\ +\ H.c.\Big\}\nonumber\\
&+& \bar{N}_\alpha \gamma_\mu (i \mbox{Im} C_{\alpha\beta}\ -\ \gamma_5
\mbox{Re}C_{\alpha\beta} )N_\beta \Bigg].
\eea 
$B$ and $C$ in Eqs.~(1) and (2) are $n_G\times 2n_G$ and $2n_G\times 2n_G$
dimensional matrices, respectively, which obey a  number of useful
identities. More details can be found in~\cite{ZPC,APetal}. For
our purpose it is sufficient to remember that the coupling matrix
$B_{i\alpha}$ is ${\cal O} (\xi)$ while the matrix $C_{\alpha\beta}$
is ${\cal O}(\xi^2)$. It is, therefore, clear that the $Z$-mediated pair
production of heavy neutrinos are more severely suppressed compared to
the W-mediated $Nl$ production due to (i) phase-space suppression and (ii)
a smaller mixing angle.

The interaction of the Majorana neutrinos with the Higgs boson is governed
by the Lagrangian
\bea
{\cal L}^{H-\nu_M-\nu_M}_{int} &=& - \frac{g_W}{4M_W}\  H
\Bigg[ \bar{\nu}_i \Big( (m_i+m_j) \mbox{Re}C_{ij}\ +\
i\gamma_5(m_j-m_i)\mbox{Im}C_{ij})\Big)\nu_j\nonumber\\
&+& 2\  \bar{\nu}_i \Big( (m_i+m_\alpha) \mbox{Re}C_{i\alpha}\ +\
i\gamma_5(m_\alpha-m_i)\mbox{Im}C_{i\alpha})\Big)N_\alpha \nonumber\\
&+& \bar{N}_\alpha \Big( (m_\alpha+m_\beta) \mbox{Re}C_{\alpha\beta}\ +\
i\gamma_5(m_\beta-m_\alpha)\mbox{Im}C_{\alpha\beta})\Big)N_\beta \Bigg].
\eea 
where $m_\alpha\ (m_i)$ stands for the mass of the $\alpha$th ($i$th) heavy
(light) neutrino. It is clear from Eq.~(3) that the coupling of the heavy
neutrinos with the Higgs boson will be enhanced by a factor $m_\alpha/M_W$.
But a similar enhancement also works, up to a different $\gamma_5$
structure, for the couplings of these Majorana neutrinos to the
longitudinal $Z$ boson or the would-be Goldstone boson $z$ in the
Feynman-'t Hooft gauge~\cite{ZPC}.
Therefore, apart from the resonance enhancement that the production
of a heavy on--shell Higgs boson and its subsequent decay
into a pair of heavy neutrinos may introduce,
a priori there is no obvious difference in the coupling strengths of the
Higgs mediated and the $Z$-mediated processes.

The bounds on the mixing angles are given in Ref.~\cite{GBetal} using both
$LEP$ results and low-energy constraints. For definiteness, we have
have used the following upper bounds from the joint fits of~\cite{GBetal}:
\bea
(s^{\nu_e}_L)^2 \ &<& \  0.01\ , \\
(s^{\nu_\mu}_L)^2 \ &<& \  0.01\ , \\
(s^{\nu_\tau}_L)^2 \ &<&\  0.065\ .
\eea 
It should be noted that these limits are obtained under the assumption that
each lepton $e$, $\mu$ or $\tau$ couples to only one heavy neutrino with
significant strength. However, in the notation of eq. 1
we can make the following identification ~\cite{GBetal}
\beq
(s^{\nu_l}_L)^2\ \ \equiv \ \  \sum\limits_\alpha |B_{l\alpha}|^2 \ .
\eeq 
Since $\tau$ lepton identification may be rather complicated  in hadron
supercolliders, we restrict  our analysis to  $LSD$ pairs of the types:
$e^+e^+$, $e^-e^-$, $\mu^+ \mu^+$, $\mu^-\mu^-$, $e^+\mu^+$ and $e^-\mu^-$
and will probe the prospects of observing lepton-number violation,
after isolating the background. On the other hand, the $LSD$
signal comprising of stable leptons which
originates from equal-sign $\tau$ leptons will eventually
be diluted by the small leptonic branching ratio of $\tau$.

\subsection*{ II.2 The cross sections}

\indent

The lepton-number violating $LSD$ signal may potentially arise due to the
following processes (see Figs.~1-3):
\bea
\mbox{A.}\qquad pp\  & \to & W^\ast W^\ast\ \to \ ll\ ,\nonumber\\
\mbox{B.}\qquad pp\  &\to & W^\ast \ \to \ lN_\alpha\ ,\nonumber\\
\mbox{C.}\qquad pp\  &\to & W^\ast Z^\ast \ \to \ lN_\alpha\ ,\nonumber\\
\mbox{D.}\qquad pp\  &\to & W^\ast \gamma^\ast \ \to \ lN_\alpha\ ,\nonumber\\
\mbox{E.}\qquad pp\  &\to & Z^\ast \ \to \ N_\alpha N_\beta\ , \nonumber\\
\mbox{F.}\qquad pp\  &\to & W^\ast W^\ast \ \to \ N_\alpha N_\beta\
,\nonumber\\
\mbox{G.}\qquad pp\  &\to & Z^\ast Z^\ast \ \to \ N_\alpha N_\beta\
,\nonumber\\
\mbox{H.}\qquad pp\  &\to & \mbox{g}\mbox{g}\ \to\ H^\ast,\ Z^\ast \ \to \
N_\alpha N_\beta \ .\nonumber
\eea %
The relevant differential cross sections ($d\hat{\sigma}_A/d\hat{t}-
d\hat{\sigma}_H/d\hat{t}$) for the parton subscatterings are listed
below
\bea
\frac{d\hat{\sigma}_A}{d\hat{t}} &=& \frac{\pi \alpha_W^2 |B^2_{l\alpha}|^2}
{4\hat{s}}\  \frac{m_\alpha^2\ (m^2_\alpha-m^2_\beta)^2}{M^4_W}
\Bigg[ \frac{\hat{t}}{(\hat{t}-m^2_\alpha)(\hat{t}-m^2_\beta)}\ +\
\frac{\hat{u}}{(\hat{u}-m^2_\alpha)(\hat{u}-m^2_\beta)} \Bigg]^2,\\[0.4cm]
\frac{d\hat{\sigma}_B}{d\hat{t}} &=& \frac{\pi \alpha_W^2 |B_{l\alpha}|^2}
{12\hat{s}^2}\ \frac{\hat{t}(\hat{t}-m^2_N)}{(\hat{s} -M^2_W)^2}\ ,\\[0.4cm]
\frac{d\hat{\sigma}_C}{d\hat{t}} &=& \frac{\pi \alpha_W^2 |B_{l\beta}
C_{\beta\alpha}|^2}{2\hat{s}^2}\ \frac{m^4_N}{M^4_W}\
\Bigg[ \frac{\hat{s}-m^2_N}{m^2_N-\hat{t}}\ +\ \frac{\hat{t}(\hat{t}-
3m^2_N)}{(\hat{t}-m^2_N)^2}\Bigg] \nonumber\\
&=& {\cal O}(\xi^6)\ ,\\[0.4cm]
\frac{d\hat{\sigma}_D}{d\hat{t}} &=& \frac{\pi \alpha_W\alpha_{em}
|B_{l\alpha}|^2}{2\hat{s}^2}\ \frac{m^2_N}{M^2_W}\ \Bigg(
\ -1\ +\ \frac{m^2_N}{\hat{s}}\ -\ \frac{\hat{s}-m^2_N}{\hat{t}}\ \Bigg),
\\[.4cm]
\frac{d\hat{\sigma}_E}{d\hat{t}}&=&\frac{\pi \alpha_W^2|C_{\alpha\beta}|^2}
{24c^4_W\hat{s}^2}\ \frac{(g^q_V)^2+(g^q_A)^2}{(\hat{s}-M^2_Z)^2}\
\Big[ (\hat{s}+\hat{t}-m^2_N)^2\ +\ (\hat{t}-m^2_N)^2\ -\ 2m^2_N\hat{s}\
\Big]\ ,\\[0.4cm]
\frac{d\hat{\sigma}_F}{d\hat{t}}&=& \frac{\pi \alpha_W^2|C_{\alpha\beta}|^2}
{2\hat{s}^2}\ \frac{m^4_N}{M^4_W}\ \Bigg[ \frac{M^2_H}{m^2_N}\
\frac{M^2_H(\hat{s}-4m^2_N)}{(\hat{s}-M^2_H)^2+M^2_H\Gamma_H^2}\
+\ \frac{\hat{s}(\hat{s}-2m^2_N)-4m^4_N}{2\hat{u}\hat{t}} \nonumber\\
&-& 1\ -\ \frac{1}{2}\ \left( \
\frac{m^4_N}{\hat{t}^2}\ +\ \frac{m^4_N}{\hat{u}^2}\ \right)\nonumber\\
&+& \frac{2M^2_H(\hat{s}-M^2_H)}{(\hat{s}-M^2_H)^2+M^2_H\Gamma_H^2}\
\left( \frac{m^2_N(\hat{s}-2m^2_N)}{\hat{u}\hat{t}}\ -\ 2 \ \right) \Bigg]
,\\[0.4cm]
\frac{d\hat{\sigma}_G}{d\hat{t}} &=& \frac{\pi \alpha_W^2|C_{\alpha\beta}|^2}
{2\hat{s}^2}\ \frac{m^4_N}{M^4_W} \ \Bigg[\ \frac{M^2_H}{m^2_N}\
\frac{M^2_H(\hat{s}-4m^2_N)}{(\hat{s}-M^2_H)^2+M^2_H\Gamma_H^2}\
+\ \frac{(\hat{s}-4m^2_N)^2}{4\hat{u}\hat{t}} \nonumber\\
&-& \left(\ \frac{m^2_N(\hat{s}-2m^2_N)}{2\hat{u}\hat{t}}\ -\ 1\ \right)^2
\Bigg], \\[0.4cm]
\frac{d\hat{\sigma}_H}{d\hat{t}} &=& \frac{\alpha_S^2\alpha_W^2
|C_{\alpha\beta}|^2}{1152\pi\hat{s}}\ \frac{m^2_N}{M^4_W}\
\left( |F^H(\frac{m^2_t}{\hat{s}})|^2
\frac{\hat{s}(\hat{s}-4m^2_N)}{(\hat{s}-M^2_H)^2+M^2_H\Gamma_H^2}\ +\
\frac{9}{4}\ |F^Z(\frac{m^2_t}{\hat{s}})|^2  \right),
\eea 
with
\bmath
F^H(x) = 3x\Big[\ 2\ +\ (4x-1)K^H(x)\ \Big],
\emath
\bmath
K^H(x) = \theta (1-4x) \frac{1}{2}\left[ \ln\left(
\frac{1+\sqrt{1-4x}}{1-\sqrt{1-4x}}\right) \ +\ i\pi \right]^2\ -
\ \theta (4x-1) 2 \left[ \sin^{-1} \left( \frac{1}{2\sqrt{x}} \right)
\right]^2,
\emath
and
\bmath
F^Z(x) = \ -(-1)^{T^q_z+1/2}\ K^Z(x),
\emath
\bea
K^Z(x) &=&  \theta(1-4x) 4x\left\{\left[ \cosh^{-1} \left( \frac{1}{2\sqrt{x}}
\right) \right]^2\ -\ \frac{\pi^2}{4}\ +\ i\pi
\cosh^{-1} \left( \frac{1}{2\sqrt{x}} \right)
\right\}\nonumber\\
&-& \theta(4x-1) 4x \left[ \sin^{-1} \left( \frac{1}{2\sqrt{x}}
\right) \right]^2. \nonumber
\eea
In Eqs.~(8)--(15), $\hat{s}$, $\hat{t}$, $\hat{u}$ are the relevant
Mandelstam variables defined at the subprocess level, $\Gamma_H$
is the total width of the Higgs boson, and
$g^q_V=-T^q_z+2Q_qs^2_W$, $g^q_A=-T^q_z$, where the third component of the
weak isospin $T^q_z$ of the $u$($d$)-type quarks and the corresponding
electric charge of them $Q_q$ (in units of $|e_{em}|$)
are respectively given by $T^{u(d)}_z=+(-)1/2$ and $Q_{u(d)}=2/3(-1/3)$.
Furthermore, Eqs.~(8),  (10),(11), (13), (14) have been computed
using  the equivalence theorem.
This simplification occurs at high energies (i.e.~$\sqrt{\hat{s}}\gg M_W$)
where one is allowed to substitute the vector bosons $W_L$ and
$Z_L$ by the corresponding would-be Goldstone bosons $w$ and $z$ in the
Landau gauge and take the limit $g_W\to 0$ by keeping $g_W/2M_W=1/v$ fixed.
This approach shown in Figs.~1--3 gives reliable results for heavy fermions
with masses $m_N\gg M_W$~\cite{DW}.
In the context of three generation models, one can further simplify the
calculations by assuming that the mass difference of each pair of heavy
neutrinos, e.g.~$N_\alpha$ and $N_\beta$, is very small compared to
the masses $m_\alpha$ and $m_\beta$, i.e. $m_\alpha,\ m_\beta \sim m_N$.
The above approximation has explicitly been employed in Eqs.~(9)--(15).

We have calculated the cross sections for the
positively charged $LSD$ pairs arising from the
$pp$ process by using
the parton distribution functions of Ref.~\cite{MRS}, $m_t=150$~GeV
and $M_H=200-1000$~GeV. The heavy neutrino masses are kept as free
phenomenological parameters. Then, the total cross sections for
the processes (B)--(C) given above are evaluated by using the generic
formula
\beq
\sigma_{tot}(l^+l^+)\ =\ \frac{1}{9}\ R^{(1)}\ \sum\limits_{ab}
\  \int dx_1 dx_2 \ f^p_a(x_1)f^p_b(x_2)\  \int d\hat{t}\
\frac{d\hat{\sigma}^0}{d\hat{t}}\ \frac{\int d\Gamma(N_\alpha \to
l^+ q \bar{q}')}{\Gamma( N_\alpha \to l q\bar{q}')},
\eeq 
where $\hat{\sigma}^0=\hat{\sigma}/|B_{l\alpha}|^2$ and
\beq
R^{(1)}\ =\
\sum\limits_{l_il_j=e,\mu}\
\sum\limits_{\alpha}\  \frac{\displaystyle |B_{l_i\alpha}|^2|B_{l_j\alpha}|^2}
{\displaystyle \sum\limits_{l_m} |B_{l_m\alpha}|^2}.
\eeq 
In models with three families, one can use the identity
that $C_{\alpha\alpha}=\sum_l |B_{l\alpha}|^2$ and the fact that
$|B_{\tau\alpha}|^2/C_{\alpha\alpha}\leq 1$ to obtain a reasonable upper bound
of
\beq
R^{(1)}_{3G} \ \ \leq\ \ (s^{\nu_e}_L)^2\  +\
(s^{\nu_\mu}_L)^2 \ ,
\eeq 
where the subscript $3G$ denotes three generations.

For the processes~(E)--(H) one uses the more involved convoluting integral
similar to Eq.~(16)
\bea
\sigma_{tot}& = &\frac{4}{81}\ R^{(2)}\ \sum\limits_{ab}
\  \int dx_1 dx_2 \ f^p_a(x_1) f^p_b(x_2)\ \int d\hat{t}
\frac{d\hat{\sigma}^0}{d\hat{t}}\
\frac{\int d\Gamma(N_\alpha \to l_i q_1 \bar{q}_1')}
{\Gamma( N_\alpha \to l_i q_1\bar{q}_2')}\nonumber\\
&&\times\ \frac{\int d\Gamma(N_\beta \to l_j q_2 \bar{q}_2')}
{\Gamma( N_\beta \to l_j q_2\bar{q}_2')},
\eea
where $\hat{\sigma}^0=\hat{\sigma}/|C_{\alpha\beta}|^2$ and
\beq
R^{(2)}\ =\
\sum\limits_{l_il_j=e,\mu}\
\sum\limits_{\alpha\beta}\  \frac{\displaystyle |B_{l_i\alpha}|^2
|C_{\alpha\beta}|^2 |B_{l_j\beta}|^2}
{\displaystyle \sum\limits_{l_ml_k} |B_{l_m\alpha}|^2 |B_{l_k\beta}|^2}.
\eeq 
Eq.\ 19 is only valid if $LSD$'s of both charges are considered.
Using similar assumptions and Schwartz's inequality,
i.e.~$C_{\alpha\alpha} C_{\beta\beta} \ge |C_{\alpha\beta}|^2$, one arrives
at the simple result
\beq
R^{(2)}_{3G} \ \leq\
\Big((s^{\nu_e}_L)^2\  +\ (s^{\nu_\mu}_L)^2\Big)^2.
\eeq 

The processes (A), (C), (D), (F) and (G) have been computed by using the
effective vector boson approximation ($EVBA$)~\cite{SD}. As we are
interested in producing heavy neutrinos with masses $m_N\ge 200 -300$~GeV,
being equivalent with a threshold invariant mass of
$\sqrt{\hat{s}_{thr}} \ge 400-500$~GeV (without including kinematical cuts
relevant for the $SM$ background), it has been demonstrated in~\cite{JOT}
that $EVBA$ can safely be applied by only using the distribution functions
of the longitudinal vector bosons. Furthermore, adapting the numerical
results of~\cite{RK}, one can readily see that the subreaction
$W_L\gamma \to lN_\alpha$ will dominate for large fermion masses ($m_N\ge
200$ GeV) by a factor of 10 at least against other subprocesses of
the type, e.g., $W_L Z_T,\ W_T Z_L,\ W_T Z_T \to l N_\alpha$ etc.

Our results are summarized in Table I. In consistency with what has been
discussed before, we find from this table that only
processes~(B) and~(D) can have sizable cross sections, i.e.~sufficiently large
to yield observable $LSD$ signals at $LHC$ or $SSC$.
The process~(A)~\cite{DKR}, though free from background sources,
is, however, suppressed by  an additional factor
$R^{(1)2}_{3G} \simeq 10^{-4}$. In the next subsection
we shall calculate the $LSD$ cross sections and compare them with the
$SM$ background.

\subsection*{II.3 The {\em LSD} signal from {\boldmath $pp \to W^\ast \to
lN_\alpha$} and the {\em SM} background}

\indent

{}From Table~I one easily concludes that the dominant contribution to the
$LSD$ signal comes from $pp\to W^\ast \to lN_\alpha$ and $pp\to
W^\ast\gamma^\ast \to lN_\alpha$. However, the cross sections for the
latter process is based on the $EVBA$. Being conservative we have not
included this process in our analysis . The numerical estimate
presented in Table~I for this process
indicates that this exclusion is not likely to alter our
conclusions at the order of magnitude level.

As has already been discussed in Ref.~\cite{DGR}, the dominant $SM$
background arises from the $t\bar{t}$ production:
\beq
pp\ \to \ t\bar{t} \ \to\ (bl^+\nu_l) (\bar{b} q_iq_j)\ ;\qquad
\bar{b}\ \to \ l^+ \nu_l c\ .
\eeq 
where $q_i q_j$ are the quarks $u$, $d$, $s$ or $c$ in appropriate
combinations. It is also important to notice that the background from
$c\bar{c}$, $b\bar{b}$ pairs or from $B^0-\bar{B}^0$ mixing will be
more severely suppressed by the lepton isolation cut (see Ref.~\cite{DGR}
for more details). We have, however, updated the analysis of Ref.~\cite{DGR}
by using the parton density functions of Ref.~\cite{MRS}.

The signal can, in principle, be distinguished from the background by the
following criteria:\\
i) The characteristics of the dilepton pairs ($p_T$ distribution, invariant
mass etc.)\\
ii) The characteristics of the jets in the final state. For example, at the
parton level the number of jets in the final state is two (four) for the signal
(background). Any conclusion based on this without taking jet fragmentation
etc.~into account, however, may turn out to be misleading. Since all
calculations in this work are based on a parton level Monte Carlo, we shall
not use the specific features of the jets. \\
iii) The signal involves only visible energy while the background has missing
$p_T$ due to the presence of stable neutrinos in the final state.
However, the missing $p_T$ spectrum (see Fig.~4), as expected, is not very
hard due to the neutrinos arising from $b$ decay. To what extent this
missing $p_T$ can be utilized in distinguishing the signal from the
background, depends crucially on the accuracy in measuring the total $p_T$
in the final state. There is no clear information on this point
 at the moment.

We have, therefore, based our analysis of improving the signal to
background ratio by solely using the characteristics of the dilepton pairs.
In any case, the simultaneous exploitation of all the three kinematical
features
listed above can only strengthen our conservative conclusions regarding the
feasibility of observing the lepton number violation at hadron colliders.

As is well known the small mass of the bottom quark relative to the large
$p_T$ of the decay lepton ensures that the lepton emerges together with the
decay hadrons within a narrow cone~\cite{GOD}, while the leptons arising from
the semileptonic decay of the heavy neutrino or the top quark are well
isolated. Hence, the background coming from the decay sequence in Eq.~(22)
can be suppressed by a suitable lepton isolation criterion
\beq
E^T_{AC}\ \ <\ \ 10\ \mbox{GeV},
\eeq 
applied to both leptons appearing in the final state. Here $E^T_{AC}$
represents the total transverse energy accompanying the lepton track within
a narrow cone of half angle 0.4 radian.

Fig.~5 shows the signal and the background cross sections against $p_{T2}$,
the transverse momentum of the softer lepton. In addition to the above
isolation cut, $p_T$ cuts $p_{T2}> 20$~GeV and $p_{T1} > 40$~GeV has been
applied, where $p_{T1}$ is the transverse momentum of the harder lepton.

It was pointed out in~\cite{DPR} that the isolation cut becomes more
effective with increasing $p_{T2}$. This is reflected in Fig.~(5) where
the background cross section goes down drastically by increasing the
$p_{T2}$ cut. It was, however, observed in Ref.~\cite{DGR} that this
dramatic reduction (obtained from a parton level Monte Carlo) may not be
completely realistic due to effects like jet fragmentation. A detailed
study of the combined effect of the lepton $p_T$ cut and the isolation cut
on the background using the $ISAJET$ program~\cite{PP} was carried out in
Ref.~\cite{MR}. The main result of Ref.~\cite{MR} was that for
the isolation cut of $E^T_{AC} < 10$~GeV, a kinematical cut $p_{T2}
> 80$~GeV suffices to kill the background completely.

Since the background can be eliminated, the prospect of detecting lepton-number
violation at hadron colliders is essentially governed by the size of the
$LSD$ signal. This signal crucially depends on the magnitude of the
mixing-angle quantity $R^{(1)}_{3G}$ and $m_N$. Using the kinematical cuts
$E^T_{AC} < 10$~GeV, $p_{T2} > 80$~GeV, the present conservative
upper bound on
$R^{(1)}_{3G} \simeq 0.02$ and an integrated luminosity
$4\times 10^5$~pb$^{-1}$/yr for $LHC$, we obtain the following results:
\beq
\barr{cc}
\qquad\qquad \underline{m_N \ [\mbox{GeV}]} \qquad\qquad & \qquad\qquad
                                             \underline{\mbox{No of $LSD$'s}}
\\
\qquad\qquad 200 \qquad\qquad & \qquad\qquad 48 \\
\qquad\qquad 300 \qquad\qquad & \qquad\qquad 32  \\
\qquad\qquad 400 \qquad\qquad & \qquad\qquad 16 \earr
\eeq 
If $LSD$'s of both signs are considered the numbers in the left column
will be multiplied by a factor of $1.5$ (approximately).
We remind the reader that in reality signals larger than the above
conservative estimates may be obtained
if a) $|B_{l\alpha}|^2$ happens to be
somewhat larger; as has already been mentioned, this possibility
is not totally excluded by the data,
if the possibility of accidental
cancellations is taken into account~\cite{LL,GBetal} b) contributions from
$pp \to W^\ast \gamma^\ast \to lN_\alpha$ are included (a detailed
calculation without using $EVBA$ is, however, desirable) c) the kinematical
cuts used in computing the cross sections can be somewhat relaxed by
exploiting other characteristics (see (ii) and (iii) above) in separating
the signal from the background. On the other hand, should $|B_{l\alpha}|^2$
(and  $R^{(1)}_{3G}$) happen to be much smaller than the existing bound,
the $LSD$ signal may remain elusive at hadron colliders.

The situation at $SSC$, however, is inconclusive at the moment. The cross
sections  happen to be larger typically by a factor 2 to 2.5 for
parameters and kinematical cuts as given above. This enhancement is not
adequate to compensate for the much smaller integrated luminosity
($10^4$~pb$^{-1}$/yr). Detailed analysis of all the avenues for enhancing
the signal as listed above is, therefore, called for. In any case, this is
also desirable in order to assess the feasibility of probing larger
neutrino masses at the $LHC$.

\section*{III.~A four-generation model with heavy Majorana neutrinos}

\subsection*{III.1~The model}

\indent

It was emphasized in~\cite{DP} that the model in Ref.~\cite{HP} predicts
large couplings of heavy Majorana neutrinos belonging to the fourth
generation with $Z$ and $H$ bosons. Hence, the production cross section
of these neutrinos, here-after denoted by $\nu$ and $N$, are expected to be
rather large at hadron colliders.

The Hill-Paschos scenario\footnote[1]{This scenario also predicts Majoron
fields, whose couplings to fermions may violate astrophysical
constraints~\cite{MZ}. However, if $m_{M_{ij}}$ are bare mass terms in the
Lagrangian or the gauge group of the $SM$ is extended by an extra
hypercharge group, $U(1)_{Y'}$, Majorons will be completely absent
in the theory.}~\cite{HP} is based on the assumption that the
$4\times 4$ mass matrices $m_D$ and $m_M$ are simultaneously diagonalizable
and $m_M=M_0${\bf 1} lies at the electroweak scale, i.e.~$0.1-1$~TeV.
Then, instead of considering  a $6\times 6$ mass
matrix, one is left with a $2\times 2 $ matrix of the form
\beq
M^\nu_i\ =\ \left( \barr{cc} 0 & m_{D_i} \\ m_{D_i} & M_0 \earr \right)\ ,
\eeq 
where the index $i$ runs over all generations. It is obvious that this
scenario corresponds effectively to an one-generation model, where the
other generations are replicas. Of course, the heavy neutrino masses
referring to the fourth generation, which are given by
\beq
m_\nu\ (m_N)\ \ =\ \frac{1}{2}\ (\sqrt{M_0^2+4m_{D_4}}\ -(+)\ M_0)\ ,
\eeq 
should have a mass larger than $M_Z/2$ in order  to be consistent with the
$LEP$ data  on neutrino-counting experiments.
This can easily be achieved if the naturalness condition
$m_{D_i}= \varepsilon_i m_{l_i}$ is assumed (motivated also by certain
$GUT$ scenarios~\cite{GUT}), where $m_{l_i}$ is the mass of the $i$th
charged lepton and the constant $\varepsilon_i$ is ${\cal O} (1)$. For the
first three generations $m_{D_i} \ll M_0$ and the light neutrinos
do not violate the experimental upper bounds~\cite{PDG}. Nevertheless, the
situation becomes different for the fourth generation. The fourth
charged lepton $E$ should be rather heavy for phenomenological reasons and
may have a mass $m_E$ ($\simeq m_{D_4}$) comparable to $M_0$. Then, both the
neutrinos belonging to this generation, i.e.~$\nu$ and $N$, can
be quite heavy so as to naturally escape detection at $LEP$ experiments.

Since the lepton mixings can effectively be recovered from the case $n_G=1$,
one has simply to make the following replacements in the differential
cross-sections given by Eqs.~(8)--(15):
\beq
B_{l\alpha} \ \to \ B_{l\nu}\ \mbox{or}\ B_{lN} \qquad \mbox{and} \qquad
C_{\alpha\beta}\ \to \ C_{\nu\nu} \ \mbox{or}\ C_{NN}.
\eeq 
Furthermore, the mixings $C_{\nu\nu}$ and $C_{NN}$ are related with the
physical heavy neutrino masses as follows:
\beq
C_{\nu\nu}\ =\ \frac{m_N}{m_\nu+m_N}\ , \qquad\quad
C_{NN}\ =\ \frac{m_\nu}{m_\nu+m_N}.
\eeq 
Finally, contributions of 4th generation quarks
to the loop functions $F^Z$ and $F^H$ in Eq.~(15) should also be considered.
Moreover, the possibility of a rather significant modification of
$\Gamma_H$ due to additional decay channels that can open should be taken
into account in the production process (H).

\subsection*{III.2 The {\em LSD} signal and the background analysis}

\indent

The $LSD$ cross sections in this model crucially depends on the relative
magnitudes of $m_E$, $m_\nu$ and $m_N$. Accordingly one can consider three
different possibilities but the dominant contribution to the $LSD$ signal
arises from the single or pair production of the $\nu$'s and especially
if $m_\nu < m_E$.

After making the replacements
as pointed out earlier in Section III.1, one finds for the cross section of
producing positively and negatively charged $LSD$'s from the process (B)
that
\beq
R^{(1)}_{4G}\ =\  \frac{(|B_{e\nu}|^2\ +\ |B_{\mu\nu}|^2)^2}
{|B_{e\nu}|^2\ +\ |B_{\mu\nu}|^2\ +\ |B_{\tau\nu}|^2}\ \leq
\ (s^{\nu_e}_L)^2\  +\ (s^{\nu_\mu}_L)^2\ .
\eeq 
Thus, we can readily conclude that an analysis similar to Sections II.2
and II.3 should apply to this case and therefore we do not intend to
repeat here, too. This also tells us that $LSD$ signals coming from the
$W$-mediated process cannot definitely address the question about the number of
neutrino generations.

We next consider the $LSD$ signal from $\nu$-pair production. The dominant
process will be the reaction (H). As already discussed in~\cite{DP},
the reason is that the quark-annihilation scattering is $\hat{s}$-channel
suppressed relative to (H). On the other hand, the presence of heavy quarks
in the triangle graph g$-$g$-H$ enhances coherently the Higgs-exchange
cross-section by a factor of 9 if all three heavy quarks are degenerate.
Since the fourth-generation up-type quark
$T$ and the corresponding down-type one $B$ should almost have equal masses
because of constraints resulting from the $\rho$-parameter or from electroweak
oblique parameters~\cite{PT},
the contribution of this additional weak isodoublet to the loop function
$F^Z$ will generally be small.

The relevant parameter $R^{(2)}$ defined in Eq.~(20) turns out to be
\beq
R^{(2)}_{4G}\ =\
|C_{\nu\nu}|^2\ \frac{(|B_{e\nu}|^2\ +\ |B_{\mu\nu}|^2)^2}
{(|B_{e\nu}|^2\ +\ |B_{\mu\nu}|^2\ +\ |B_{\tau\nu}|^2)^2}\ \leq
\ \frac{m^2_N}{(m_\nu+m_N)^2}.
\eeq 
In fact, there is no strong upper bound on the parameter
$R^{(2)}_{4G}$, which can approach the unity for $m_N\gg m_\nu$~\cite{BS}.
This is a quite remarkable
observation if one compares with numerical results presented in Table I
for three-generation models
which are suppressed by an additional lepton-violating-mixing factor
$(0.02)^2 = 4.\ 10^{-4}$.

As an illustration we have considered the following values for the
parameters: $M_0=100$~GeV, $m_E=320$~GeV and $\varepsilon=0.75$. This
yields $m_\nu=195$~GeV and $m_N=295$~GeV. We then compute the $LSD$ cross
section (including like-sign $e$ and $\mu$'s of both charges in the final
state) subject to the kinematical cuts on the leptons discussed in Section
III.3 which suffice to remove the background from $t\bar{t}$ production.
We have also taken $m_T\simeq m_B = 400$~GeV. The additional decay modes
of the Higgs boson leading to a modification of $\Gamma_H$, as discussed
above, have also been taken into account. The results for $LHC$ ($SSC$)
 energies for various Higgs masses ($M_H$) are displayed in Table IIa (IIb).
\centerline{{\bf  Table IIa}}
\beq
\barr{cc}
\qquad\qquad \underline{M_H~[\mbox{GeV}]} \qquad\qquad &
                      \qquad\qquad \underline{\mbox{No.~of events/year}}\\
\ 200 & \qquad\qquad \ 8600\times R^{(2)}_{4G}\\
\ 400 & \qquad\qquad 13300\times R^{(2)}_{4G}\\
\ 600 & \qquad\qquad 34000 \times R^{(2)}_{4G}\\
\ 800 & \qquad\qquad 17300 \times R^{(2)}_{4G}\\
1000  & \qquad\qquad \ 7200 \times R^{(2)}_{4G}\earr
\eeq 
\centerline{{\bf  Table IIb}}
\beq
\barr{cc}
\qquad\qquad \underline{M_H~[\mbox{GeV}]} \qquad\qquad &
                      \qquad\qquad \underline{\mbox{No.~of events/year}}\\
\ 200 & \qquad\qquad \ 1150\times R^{(2)}_{4G}\\
\ 400 & \qquad\qquad \ 2000\times R^{(2)}_{4G}\\
\ 600 & \qquad\qquad \ 4200 \times R^{(2)}_{4G}\\
\ 800 & \qquad\qquad \ 2600 \times R^{(2)}_{4G}\\
1000  & \qquad\qquad \ 1050 \times R^{(2)}_{4G}\earr
\eeq
Thus, even with $R^{(2)}_{4G} \simeq 10^{-3}$ a reasonable number
of background-free events may be expected at $LHC$. At $SSC$, on the other
hand, a value of $R^{(2)}_{4G} \simeq 10^{-2}$ may yield an
observable $LSD$ signal. The enhancement due to the on-shell production of
the Higgs boson and its subsequent decay into $\nu$ pairs, as discussed in
Ref.~\cite{DP}, can be traced back from the above tables.

In principle, the background arising due to
 $LSD$ pairs from $T\bar{T}$, $B\bar{B}$
production followed by cascade decays similar to Eq.~(22) should also be
considered. In the absence of any information about the partial decay rates
of $T$ and $B$ a complete analysis cannot be made. However, the following
arguments will convince the reader that a substantial background from this
channel is not likely to occur:\\
i) The production cross section of $T\bar{T}$ and $B\bar{B}$ are much
suppressed compared to $t\bar{t}$ production. For example, with $m_t=
150$~GeV, $m_T\simeq m_B \simeq 400$~GeV, we have estimated  that
$\sigma_{T\bar{T}}/\sigma_{t\bar{t}} \simeq 10^{-2}$ at $LHC$ energies.\\
ii) As a plausible scenario we have assumed that Br$(T \to B+X)\simeq 1$
and Br$(B\to t+X)\simeq 1$. The $LSD$ signal may then arise through the
decay chains
\bmath
\barr{c}
T \ \to \ Bl^+\nu_l \qquad \mbox{and}\\
\bar{T} \ \to \ \bar{B} X; \qquad \bar{B}\ \to\ \bar{t} l^+\nu_l \earr
\emath
Since the mass difference between $T$ and $B$ cannot be very large for
reasons mentioned above, simple decay kinematics will indicate that
both the leptons are soft and are not likely to survive the stringent
$p_T$ cuts which are, in any case, required to eliminate the background
from $t\bar{t}$ pairs. Our Monte-Carlo calculations using
$m_T=400$~GeV, $m_B=360$~GeV and $m_t=150$~GeV and kinematical cuts as
discussed in section II supports this conclusion.

\section*{IV.~Conclusions}

\indent

In this paper we have studied both three and four generation models
with heavy Majorana neutrinos, based on the gauge group
$SU(2)_L \otimes U(1)_Y$. We have computed all possible cross--sections
for the production of such neutrinos, either singly or in pairs,
using parton level Monte Carlo. Our calculations reveal that in the three
generation model the dominant cross--section is given by the processes
(B) and (D) of section II where heavy neutrinos are singly produced in
association with a lepton. Lepton number violation arising through the
decays of these neutrinos can be detected at the $LHC$ by looking for
high $p_{T}$ $LSD$ pairs ($p_{T} > 80$~GeV) provided certain mixing
angles are not too small compared to their existing upper bounds and
the mass of these neutrinos are $< 400$ GeV. A similar analysis reveals
that the isolation of a background free sample of dileptons at $SSC$
is not very likely by looking for high $p_{T}$ leptons only. In order to
do this or to probe larger mass ranges at $LHC$ other features of the signal
(e.g., the characteristics of the jets) should be properly utilized.
Further studies taking effects like jet fragmentations into account are,
therefore, called for.

Calculations in the four generation model reveal that the pair production
of these neutrinos through the processes (H) given in section~II may also
turn out to be the most dominant source of $LSD$'s. This cross section
is not suppressed by any small mixing angles but rather depends on the
ratio of certain mixing angles. No strong bound on this ratio exists
at the moment. Sizable background free $LSD$ samples observable at both $LHC$
and $SSC$ are predicted in this scenario.\\[2.5cm]
{\bf Acknowledgements.} One of us ($AD$) wishes to thank Prof.~K.\
Kleinknecht for the kind hospitality in the University of Mainz, FRG.
done.  $AD$ also thanks the Department of Science and Technology, Government
of India for the grant of a research project. $MG$ thanks for the grant of a
senior fellowship from the Council of Scientific and Industrial Research,
India.

\newpage

\centerline{\bf\Large Figure Captions }
\vspace{1cm}
\newcounter{fig}
\begin{list}{\bf\rm Fig. \arabic{fig}: }{\usecounter{fig}
\labelwidth1.6cm \leftmargin2.5cm \labelsep0.4cm \itemsep0ex plus0.2ex }

\item Feynman graphs responsible for the subprocess (A): $W_LW_L\to
l^+l^+$.

\item Feynman graphs relevant for the singly heavy Majorana neutrino
production, i.e. processes (B), (C) and (D) (see also text).

\item Feynman diagrams relevant for double heavy Majorana neutrino
production as described by the processes (E)--(H) in Section II.2

\item Missing transverse momentum distribution of the $SM$ background
(see also Eq.~(22)).

\item Transverse momentum distribution of the softer lepton $p_{T2}$ coming
from the $SM$ background. For comparison, we have considered the $p_{T2}$
distribution of the $LSD$ signal which predominantly originates from
the process (B).

\end{list}

\newpage

\noindent
{\bf Table I.}~Numerical
estimates of production cross sections for the processes (A)--(H)
leading to $LSD$ signals in the context of three-generation models.\\[1.cm]
\begin{tabular*}{12.34cm}{|c||c|c|}
\hline
 & &  \\
 & $m_{N}= 200 - 1000$~GeV & $m_{N}= 200 - 1000$~GeV \\
Process &$LHC\ (\sqrt{s}=16)$~TeV & $SSC\ (\sqrt{s}=16)$~TeV\\
 & $\sigma_{tot}$~[pb] & $\sigma_{tot}$~[pb] \\
\hline\hline
&& \\
A. & $<\ 5.\ 10^{-2} \times R^{(1)2}$ & $<\ 1.\ 10^{-1} \times R^{(1)2}$\\
B. & $1.\ -\ 2.\ 10^{-3} \times R^{(1)}$ & $15.\ -\ 3.\ 10^{-2} \times R^{(1)}$
                                                                           \\
C. & small, ${\cal O}(R^{(1)3})$ & small, ${\cal O}(R^{(1)3})$ \\
D. & $ 3.\ 10^{-3}\times R^{(1)}$ & $ 4.5\ 10^{-2}\times R^{(1)}$ \\
E. & $5.\ (10^{-4}-10^{-6})\times R^{(2)}$ &
                                      $ (10^{-3} - 10^{-5})\times R^{(2)}$
                                                                          \\
F. & $(2.5\ 10^{-4}- 5.\ 10^{-3})\times R^{(2)}$ &
                           $(3.\ 10^{-3}- 8.\ 10^{-2})\times R^{(2)}$  \\
G. & $(2.\ 10^{-4}- 4.\ 10^{-3})\times R^{(2)}$ &
                            $(2.5\ 10^{-3}- 7.\ 10^{-2})\times R^{(2)}$ \\
H. & $5.\ (10^{-3} - 10^{-5})\times R^{(2)}$    &
                                $4.5\ (10^{-2} - 10^{-4})\times R^{(2)}$ \\
&&\\
\hline
\end{tabular*}

\end{document}